# Neutron powder diffraction investigation of the structural and magnetic properties of $(La_{1-y}Y_y)FeAsO$

A. Martinelli[1,*], A. Palenzona[1,2], M. Tropeano[1,3], C. Ferdeghini[1], M. R. Cimberle[4], C. Ritter[5]

[1] *CNR-INFM-LAMIA Corso Perrone 24, 16152 Genova – Italy*

[2] *Dipartimento di Chimica e Chimica Industriale, Università di Genova, via Dodecaneso 31, 16146 Genova – Italy*

[3] *Dipartimento di Fisica, Università di Genova, via Dodecaneso 33, 16146 Genova – Italy*

[4] *CNR-IMEM, Dipartimento di Fisica, Via Dodecaneso 33, 16146 Genova, Italy*

[5] *Institute Laue - Langevin, 6 rue Jules Horowitz, 38042 Grenoble Cedex 9 – France*

**Abstract**

The structural, magnetic and resistive properties of $(La_{1-y}Y_y)FeAsO$ compounds ($y$ = 0.10, 0.20, 0.30) have been investigated by means of X-ray and neutron powder diffraction as well as by resistivity measurements. The temperatures at which the structural transition from tetragonal to orthorhombic and the magnetic ordering take place progressively reduce by similar amounts with increasing Y substitution on account of the progressive chemical pressure increase. We propose that the structural transition could be originated by a cooperative Jahn-Teller distortion involving the alignment of the fully occupied Fe $3d_{z^2}$ orbitals in the Fe plane along the *y* axis, leading to the branching of the cell parameters *a* and *b*. The magnetic structure develops after the occurrence of the structural transition, but before its completion.

---

[*] Corresponding author: amartin@chimica.unige.it

# 1. Introduction

The class of compounds referred to as oxy-pnictide and characterized by the general formula *RE*FeAsO (*RE*: rare earth) attracted much attention after the discovery of a relatively high superconducting transition temperature ($T_C \sim 26$ K) in LaFeAs($O_{1-x}F_x$) by Kamihara et al. [1]. *RE*FeAsO compounds (belonging to the so called 1111-family) undergo a tetragonal to orthorhombic phase transition at $T_{T-O}$ on cooling, coupled with a long-range spin density wave (SDW) type antiferromagnetic order at slightly lower temperature. If F-substitution (leading to electron doping) or O vacancies (leading to hole doping) exceeds a critical value in these compounds, the tetragonal to orthorhombic phase transition is suppressed as well as the SDW, whereas superconductivity (SC) arises.[2,3]

Within the 1111 family, La-based compounds are characterized by the lowest superconductive critical temperatures ($T_C$), probably on account of the large ionic radius (IR) of La. Hence these compounds represent one of the best systems to investigate the effects of the external and chemical pressure on structural, magnetic and superconductive properties. The role of the chemical pressure on the superconductive properties was investigated in O-deficient $(La_{1-y}Y_y)FeAsO_{1-x}$ compounds and $T_C$ up to 43 K was detected for optimally hole doped samples with $y = 0.40$.[4] A similar positive effect of the chemical pressure on superconductive properties was also observed in hole doped $(La_{1-y}Sm_y)FeAsO_{0.85}$.[5] The effect of external pressure on the structural transition and/or SDW in undoped compounds was investigated in several experiments. A decrease of $T_{SDW}$ with pressure was observed in a slightly F-doped SmFeAsO sample,[6] whereas a complete suppression of magnetism was obtained in $CaFe_2As_2$ and $(K_xSr_{1-x})Fe_2As_2$ compounds,[7,8] belonging to the so called 122-family. In a previous investigation[9] we have demonstrated that both $T_C$ in $(La_{1-y}Y_y)FeAs(O_{0.85}F_{0.15})$ samples and $T_{SDW}$ in the parent $(La_{1-y}Y_y)FeAsO$ samples present a clear dependence on the Y content. In particular it was shown that Y substitution at the La site is effective in producing chemical pressure, with $T_{SDW}$ decreasing with the increase of the Y content; conversely in the F doped samples $T_C$ increases.

In this work the effect of the progressive reduction of the IR values and the disorder at the *RE* site on both the structural transition and the occurrence of the SDW is investigated.

**2. Experimental**

Samples of the series $(La_{1-y}Y_y)FeAsO$, with $y$ = 0.10, 0.20, 0.30, 0.50, were synthesized by a solid state reaction using a four step reaction procedure, slightly different from that previously applied[10] for the preparation of SmFeAsO: 1) synthesis of the intermetallic solid solution $La_{1-y}Y_y$; 2) reaction of $La_{1-y}Y_y$ with stoichiometric amount of As in order to obtain $(La_{1-y}Y_y)As$; 3) synthesis of $(La_{1-y}Y_y)FeAsO$ by reacting stoichiometric amounts of $(La_{1-y}Y_y)As$ with Fe and $Fe_2O_3$ in a form of a pellet in an evacuated pyrex flask at 500°C for 1-2 days; 4) the so obtained product was ground, well mixed and pressed in a new pellet and heated in an evacuated quartz flask at 1100°C for 70 h. All the handling and manipulation of the samples was carried out in a glove box where the working atmosphere was continuously purified to values less than 1 ppm with regard to $H_2O/O_2$. Phase identification was performed by X-ray powder diffraction at room temperature (XRPD; PHILIPS PW3020; Bragg-Brentano geometry; $CuK_{\alpha1, \alpha2}$; range 15 – 120° $2q$ ; step 0.020° $2q$; sampling time 10 s). Neutron powder diffraction (NPD) analysis was carried out at the Institute Laue Langevin (Grenoble – France) for the samples with $y$ = 0.10, 0.20, 0.30. In order to evaluate the temperature at which the structural transition takes place thermo-diffractograms were acquired on heating in continuous scanning mode in the $T$ range 2 – 200 K using the high intensity D1B diffractometer ($\lambda$ = 2.52 Å). High resolution NPD patterns were collected at selected temperatures between 10 K and 300 K using the D1A diffractometer ($\lambda$ = 1.91 Å). Rietveld refinement of both XRPD and NPD data was carried out using the program FULLPROF;[11] by means of a $LaB_6$ (XRPD) or NAC (NPD) standard an instrumental resolution file was obtained and applied during refinements in order to detect micro-structural contribution to both XRPD and NPD peak shape. The diffraction lines were modelled by a Thompson-Cox-Hastings pseudo-Voigt convoluted with axial divergence asymmetry function[12] and the background by a fifth-order polynomial. The following parameters were refined in the final refinement cycle: the overall scale factor; the background (five parameters of the $5^{th}$ order polynomial); $2q$-Zero; the unit cell parameters; the specimen displacement; the reflection-profile asymmetry; the $z$ Wyckoff positions of (La,Y) and O (both located at the $2c$ site); the isotropic thermal parameters $B$; the anisotropic strain parameters.

Resistive measurements were carried out using a standard four probe technique in the range $T$ 5-300 K.

## 3. Results

XRPD patterns reveal the formation of $(La_{1-y}Y_y)FeAsO$ in all cases; little amounts of $Y_2O_3$ are present in the samples with $y \leq 0.30$, whereas for the sample with $y = 0.50$ the amount of the secondary oxide is notably increased, probably on account of internal strains (see below) that decrease the thermodynamic stability of the oxy-pnictide.

The iso-valent substitution of La with Y determines a progressive decrease of the average IR size at the *RE* site (<IR> = 1.146 Å, 1.132 Å, 1.118 Å, 1.089 Å for $y$ = 0.10, 0.20, 0.30 and 0.50, respectively)[13] so that $a$ and $c$ axes decrease with the increase of Y content by the same amount in percentage, as revealed by the Rietveld refinement of XRPD data (Table 1). The value of the $z$ coordinate at the *RE* site decreases with the increase of Y content; conversely the $z$ coordinate at the As site increases. In particular the samples with $y$ = 0.10 and 0.20 share about the same values, within the experimental error, whereas a marked deviation is exhibited by the sample with $y$ = 0.30 and 0.50. As a result in the former two samples the distance between neighbouring $[(La,Y)O]^{+1}$ and $[FeAs]^{-1}$ tetrahedral layers is about the same (~ 0.208 Å), whereas in the latter ones this distance reduces to ~ 0.207 Å and ~ 0.206 Å for $y$ = 0.30 and 0.50, respectively (in Figure 1 the crystal structure is drawn). The Fe-As bond lengths decrease with Y substitution, thus indicating an increasing compressing strain at the Fe site, whereas the tetrahedral distortion, measured by the As-Fe-As bond angle, remains almost constant (Table 1).

Information on the micro-structural strains occurring within the samples was obtained analyzing the broadening of XRPD lines by means of the Williamson-Hall plot method.[14] Generally, in the case that size effects are negligible, a straight line passing through the origin has to be observed, whereas the slope provides the lattice strain. When broadening is not isotropic, size and strain effects along some crystallographic directions can be obtained by considering different orders of the same reflection. In our case, for each sample, size contribution is negligible, since a straight line passing through the origin can be traced, and micro-strains progressively increase as the crystallographic direction approaches [001]. The comparison of the Williamson-Hall plots reveals that lattice strains increase with the Y content; for clarity Figure 2 shows only the Williamson-Hall plots for the family of 00$l$ peaks.

In order to detect the atomic sites which are most affected by strains, the bond valence sum (BVS) method[15] can be applied; by means of this method bond valences are calculated from the experimental bond lengths and the sum of the bond valences around any atom is equal to its formal valence state. In general the difference between the calculated BVS and the atomic valence is usually small (< 0.1 valence units - vu) if bond stretching or compression is negligible; hence a large deviation from the expected valence value can be indicative of strained bonds.[16] The bond valence parameter ($R_0$) for the $Fe^{2+}$-$As^{3-}$ bond is not tabulated and was estimated by applying the equation (3) reported in ref.[15] using the available crystallographic data for the polymorphic modifications of FeAs, obtaining $R_0 = 2.175$ Å. BVS values reported in Table 1 are in quite good agreement with the nominal values and reveal that the atoms located at the vertexes of the tetrahedra are not affected by strain, whereas O and Fe, both at the centres of these polyhedra, undergo a notable strain. In particular the strain at the O site decreases with the increase of Y content and this is easily explained by the fact that La is progressively substituted by a lighter atomic species such as Y. Conversely strains at Fe site increase with Y content on account of chemical pressure that progressively reduces the cell size; in fact in $(La_{1-y}Y_y)FeAsO$ three main bonds are present: (La,Y)-O, (La,Y)-As and Fe-As. It is evident that for both (La,Y)-O and (La,Y)-

As bonds the reduction of the cell size, and consequently of the bond lengths, is somehow compensated by the reduction of <IR> at the *RE* site, whereas this is not possible for the Fe-As bond, whose length decreases with the increase of the Y content (Table 1). In addition the shape of the [FeAs]$^{-1}$ tetrahedron is almost unaffected by chemical substitution, since the As-Fe-As bond angles undergo an almost negligible variation; hence the environment of Fe is not distorted and this ion, located at the centre of a tetrahedral cage, undergoes a progressive compression with the increase of Y content, as revealed by the increase of the BVS. An opposite effect is experienced by O, whose cage becomes progressively larger thanks to the reduction of <IR> at the *RE* site, despite the progressive shortening of the (La,Y)-O and (La,Y)-As bond lengths (Table 1). This result shows the effectiveness of Y substitution in applying chemical pressure on Fe site.

Thermo-diffractograms collected using D1B were used to determine the onset of both structural and magnetic transition. The 220 diffraction peak is usually analyzed to determine the tetragonal to orthorhombic ($T_{T-O}$) structural transition using NPD data, since it is quite strong and undergoes an evident splitting at the structural transition; conversely the 110 diffraction peak has almost null intensity in NPD patterns. Unfortunately the 220 peak is out of the range of the D1B diffractometer, at least in our experimental conditions; as a consequence no evident peak splitting can be observed in the D1B NPD. In any case some *hhl* peaks are present in these patterns and although they don't exhibit evident splitting, they broaden on cooling. For this reason, in order to identify the $T_{T-O}$ of the different samples, the dependence on *T* of the Lorentzian isotropic strain (LIS) contribution to the 112 diffraction peak (the strongest of these *hhl* peaks) was analyzed carrying out a peak fit procedure using an instrumental resolution file. In this way it was possible to determine the $T_{T-O}$: Figure 3 shows the evolution of the LIS with temperature for the analyzed samples. The $T_{T-O}$ have been obtained by a careful analysis of these data and are reported in Table 2 together with the magnetic and resistive transition temperatures. The $T_{T-O}$ values are in quite good agreement with the data reported for LaFeAsO, whose structural transition is located at 155-160 K,[17,18,19] revealing that Y substitution progressively decreases the $T_{T-O}$.

Figure 4 shows that the LIS of the (102+003) peak of $(La_{0.90}Y_{0.10})FeAsO$, selected as representative for peaks with $hkl \neq hhl$, is practically constant in whole inspected range, as expected, thus confirming that the applied method can be used to determine selective splitting originated by structural transition in our samples.

A careful analysis of the thermodiffractograms collected at low temperature reveals the appearance of an extremely faint contribution to scattering at ~4.01 Å; according to what had been previously observed in LaFeAsO this additional scattering comes from the magnetic ordering.[19] Figure 5 shows the dependence of the neutron scattering intensity at ~4.01 Å on temperature for the three samples: a careful analysis of the data reveals that the onset of the $T_{SDW}$ decreases with the increase of the Y content (Table 2). Previous works investigating LaFeAsO by NPD analysis reported slightly different values for $T_{SDW}$, ranging from 137 K to <145 K;[17,19] these differences can be ascribed to the faintness of the magnetic scattering contribution to the diffraction pattern, that make the exact ascertainment of the $T_{SDW}$ quite difficult.

It is interesting to observe that whatever the composition the onset of $T_{T-O}$ is always some degrees higher than $T_{SDW}$; within the experimental uncertainty the difference between these two temperatures seems to remain constant in the three samples. We can state that the magnetic ordering takes place during the structural transformation and before its completion. This result suggests on the one hand that the structural transition is not driven by magnetic interactions taking place among the $Fe^{2+}$ ions but as well that in-plane magnetic interactions occur before the completion of the orbital ordering at the Fe site which is probably driving the structural transition. In fact theoretical calculations suggest that the ordering of the Fe orbitals is responsible for the lattice distortion[20]; we will discuss this topic in detail in the Discussion.

Rietveld refinements using the D1A data were carried out using a tetragonal structural model for data collected above $T_{T-O}$ and an orthorhombic one for data collected below $T_{T-O}$. In Table 3 are reported the structural data of the three samples at 10 K, whereas Figure 6 shows the evolution of the cell parameters as a function of temperature for the three samples. Noteworthy the *a*/*b* ratio remains constant in the three samples, indicating that even though Y substitution hinders to some extent the structural transition, as revealed by the decrease of $T_{T-O}$, the same level of ordering of the Fe orbitals is obtained, whatever the composition.

Figure 7 shows the superposition of Williamson-Hall plots obtained from NPD data collected between 100 K and 200 K on the sample $(La_{0.90}Y_{0.10})FeAsO$: it is evident that as the structural transition is approached, cooling from 200 K down to 160 K, an anomalous broadening due to structural strain characterizes the *hhl* and *32l* families of peaks, that is suppressed after the structural transition. Conversely, the remaining families of peaks do not exhibit any particular anomalous behaviour in this temperature range. This behaviour reveals that structural strain in the *ab* plane of the tetragonal structure undergoes an abrupt increase just above $T_{T-O}$ that is rapidly suppressed by the structural transition. As we will discuss in the next section, this behaviour can be ascribed to the tendency of the Fe $3d_{z^2}$ orbitals to order in the Fe plane as the structural transition is approached on cooling; with the occurrence of the orthorhombic structure, orbital ordering can be accommodated by the structure and strains are suppressed.

The transition temperatures as detected by neutron diffraction can be compared with those obtained by resistivity measurements. Normalized to room temperature and shifted for a better visualization the resistivity curves of the $(La_{1-y}Y_y)FeAsO$ samples are displayed in Figure 8. They show the typical behaviour of undoped 1111 oxypnictides[7] compounds with a maximum followed by a sharp drop with an inflection point at $T_{drop}$ (corresponding to the maximum of d$r$/d$T$), related to the occurrence of the SDW. Increasing the Y content, $T_{drop}$ shifts to lower temperatures (Figure 8, inset). The values of $T_{drop}$ reported in Table 2 are in excellent agreement with the $T_{SDW}$ values obtained by NPD analysis, thus confirming the relationship between $T_{drop}$ observed in the resistivity curves and the onset of the SDW.

**4 Discussion**

Y substitution for La affects the structural and physical properties of $(La_{1-y}Y_y)FeAsO$ samples through the combination of two main effects: 1) the chemical pressure induced by the substitution of La with the smaller Y, resulting in a progressive increase of the strain at the Fe site; 2) the chemical disorder at the $RE$ site that can be measured by the variance,[21] defined as:

$$\sigma^2 = \sum_i y_i IR_i^2 - \left(\sum_i y_i IR_i\right)^2$$

where $y_i$ are the proportions of each ion present.

Our data shows that $T_{drop}$ decreases in $(La_{1-y}Y_y)FeAsO$ with the increase of the Y content. Taking into account the data reported in ref.[9] it is possible to compare the $(La_{0.70}Y_{0.30})FeAsO$ and $(La_{0.30}Y_{0.70})FeAsO$ samples, characterized by the same degree of disorder at the $RE$ site, but different level of chemical pressure. These samples are characterized by $T_{drop}$ values equal to 128 K and 120 K, respectively, indicating that chemical pressure alone hinders independent of chemical disorder the SDW and probably the tetragonal to orthorhombic transition as well. It is then expected that exerting chemical pressure by doping induces as well an increase of $T_C$. This has been experimentally observed in the homologous F-doped samples, characterized by $T_C$ values that increase with chemical pressure,[9] in agreement with the rising of $T_C$ observed by application of external pressure.[22]

In order to evaluate how chemical disorder affects the properties of these materials a series of compounds characterised by the same value of <IR> at the $RE$ site, but different variance should be compared. This experiment exceeds our work, but the $(La_{0.90}Y_{0.10})FeAsO$ sample has almost the same <IR> value at the $RE$ site as CeFeAsO (1.146 Å and 1.143 Å, respectively); CeFeAsO is characterized by a $T_{T-O}$ ~155 K[23] and a $T_{drop}$ ~145 K,[24] both transition temperatures are therefore ~3 K higher than the homologous temperatures measured in $(La_{0.90}Y_{0.10})FeAsO$.

These results suggest that both chemical disorder at the $RE$ site and chemical pressure hinder the structural transition and the magnetic ordering, as evidenced by the decrease of $T_{T-O}$, $T_{SDW}$ and $T_{drop}$.

An interesting result is that the magnetic structure develops before the completion of the structural transition, thus suggesting that the occurrence of the orthorhombic structure is not driven by magnetism. Taking into account the removal of degeneracy between Fe $3d_{x^2-y^2}$ and $3d_{z^2}$ orbitals due to the Jahn-Teller effect, it has been suggested that a competition involving super-exchange anti-ferromagnetic interactions between a pair of next-nearest-neighbour $3d_{x^2-y^2}$ Fe spins through an intervening As $4p_{x-y}$ (or $4p_{x+y}$) orbital and super-exchange ferromagnetic interactions between a pair of nearest-neighbour $3d_{x^2-y^2}$ Fe spins through a pair of distinct intervening As $4p_{x-y}$ and $4p_{x+y}$ orbitals exists.[25] This scenario demands that the Fe $3d_{x^2-y^2}$ orbitals must lie in the Fe plane, and contrasts therefore with the 2-fold axis symmetry of this plane at low temperature. In this context it is worth noting that the occurrence of the orthorhombic structure leads to the formation of two different Fe-Fe bond lengths, on account of $3d$ Fe orbital ordering. Moreover at room temperature the Jahn-Teller $Fe^{2+}$ ion is located at the $2b$ site of the $P4/nmm$ space group; hence its environment is not perfectly tetrahedral, since this site has tetragonal ($\overline{4}m2$) point group symmetry. As a consequence, in addition to the crystal field splitting of the $t_2$ and $e$ orbitals arising from the tetrahedral coordination around the $Fe^{2+}$ site, the on site tetragonal crystal field further splits the $d$ levels, thus giving rise to a Jahn-Teller distortion. It is important now too take into account two features: 1) according to ref. [25] the Fe $3d_{z^2}$ orbital is fully occupied and hence strongly negative; 2) a regular tetrahedron can be inscribed within a cube, but the tetrahedra in the *RE*FeAsO compounds are tetragonally distorted, strongly compressed along the $z$ axis with pseudo-cubic cell edges $a_c = b_c \sim 4$ Å and $c_c \sim 2.6$ Å. As a consequence the Coulomb repulsion prevents the orientation of the Fe $3d_{z^2}$ orbitals along the $c$ axis, because of the interaction with the neighbouring negative charged As $4p_x$ and $4p_y$ orbitals; conversely it is expected that this orbital lies in the $ab$ plane (the Fe plane). At room temperature Fe $3d_{z^2}$ orbitals are randomly arranged in the Fe plane; on cooling a cooperative Jahn-Teller distortion takes place leading to an orbital ordering in the Fe plane and thus to the tetragonal to orthorhombic transition (note that a cooperative Jahn-Teller effect was predicted for orthorhombic LaFeAsO).[18] In this scenario the cell parameters *a* and *b* should branch at $T_{T-O}$ due to the $3d_{z^2}$ orbital ordering (Fe-Fe bonds are parallel to *x* and *y* and their values correspond to *a*/2 and *b*/2, see Figure 6), whereas *c* should not be affected by any particular feature (the $3d_{x^2-y^2}$ orbitals, being perpendicular to the $3d_{z^2}$, ones are lying in the *ac* plane). This is exactly what is observed in Figure 6: the alignment of the Fe $3d_{z^2}$ orbitals along the *b* direction (Figure 9) induces an abrupt elongation of the Fe-Fe bonds along it, because of the Coulomb repulsion due to the full occupation of the Fe $3d_{z^2}$ orbitals, whereas a simultaneous compression along the *x* axis determines the decrease of *a*. As a result of the $3d_{z^2}$ orbital ordering, the cell parameters *a* and *b* branch. Orbital ordering can explain the evolution of the structural strain depicted in Figure 7: as the structural transition is approached on cooling, the Fe $3d_{z^2}$ orbitals tend to order in the tetragonal phase and hence strains arise in the Fe plane. At 160 K exceeding structural strain at the Fe plane gains its maximum, but soon after the occurrence of the orthorhombic structure it is completely suppressed.

Figure 6 shows that the average value of orthorhombic cell parameters *a* and *b* (representing a hypothetical disordered arrangement of the $3d_{z^2}$ orbitals in the Fe plane) follows the behaviour expected for the tetragonal cell edge *a* at low *T*, thus implying no transition in the specific volume vs *T* behaviour. This feature is typical of order-disorder phase transitions and is observed, for example, in manganites undergoing an orbital ordering due to Jahn-Teller distortion.[26]

In F-substituted *RE*FeAsO samples doping electrons partially fill the Fe $3d_{x^2-y^2}$ orbitals[25] leading to the progressive degeneracy of the Fe *e* orbitals as well as to the suppression of the cooperative Jahn-Teller distortion. As a consequence the branching of the cell parameters *a* and *b* decreases with the increase of the electron doping up to the suppression of the orbital ordering.[2,3,23]

By theoretical investigation[27] it was proposed that the structural distortion is driven by anti-ferromagnetic ordering, resulting in different occupancies of the Fe $3d_{xz}$ and $3d_{yz}$ orbitals in case of the monoclinic angle *g* being slightly higher than 90°. It was as well recognized that with *g* = 90° the Fe $3d_{xz}$ and $3d_{yz}$ orbitals would be degenerate and anti-ferromagnetic ordering could not drive the structural transition. Some experimental evidences contrast, however, with this scenario foreseeing the structural distortion driven by the anti-ferromagnetic ordering. First of all, as pointed out by McGuire *et al.*[19], there is some confusion regarding the low temperature structure of the undoped compounds: initially reported as belonging to the monoclinic system (*g* ≠ 90°),[17] it has in the meantime been recognized as orthorhombic (*g* = 90°).[2,3,18,19,23,28,29] Therefore super-exchange interactions are not the driving force for the structural transition. Secondly there is no relationship between the strength of super-exchange interactions (related to the value of the ordered magnetic moment) and $T_{SDW}$; for example LaFeAsO exhibits the highest $T_{SDW}$ as well as the lowest value of ordered magnetic moment at the Fe site. Finally several investigations demonstrated the occurrence of the orthorhombic structure at low temperature in different F-doped superconducting *RE*FeAsO compounds.[23,2,3] In these cases orbital ordering takes place, but anti-ferromagnetic ordering at the Fe site is suppressed in favour of competing superconductivity.

**Conclusions**

Samples with nominal composition $(La_{1-y}Y_y)FeAsO$ were analyzed by means of neutron powder diffraction analysis between 10 K and room temperature. The tetragonal to orthorhombic transition temperatures as well as the temperature at which the spin density wave appears were determined. The spin density wave temperature is in a very good agreement with the maximum observed in the first derivative of the resistivity curves, whatever the sample composition. Y substitution induces chemical disorder at the *RE* site and chemical pressure within the structure. Both factors influence the structural transition and the magnetic ordering, as evidenced by the decrease of $T_{T-O}$, $T_{SDW}$ and

$T_{drop}$. In the light of the neutron powder diffraction analysis we propose a scenario involving the ordering of the Fe $3d_{z^2}$ orbitals along the *b* direction as a result of a cooperative Jahn-Teller distortion, driving to the observed tetragonal to orthorhombic structural transition. This scenario is consistent with the main structural features observed in *RE*FeAsO compounds.

**Acknowledgments**

This work is partially supported by the "Compagnia di S. Paolo" and by the Italian Foreign Affairs Ministry (MAE), General Direction for the Cultural Promotion.

Table 1: Structural data at room temperature for (La$_{1-y}$Y$_y$)FeAsO samples obtained by Rietveld refinement of XRPD data (space group no. 129 - P4/*nmm*, origin choice 2; La,Y at 2*c*; Fe at 2*b*; As at 2*c*; O at 2*a*); bond valence sum (BVS) values (measured as valence units) are also reported.

|  | **y = 0.10** | **y = 0.20** | **y = 0.30** | **y = 0.50** |
|---|---|---|---|---|
| *a* (Å) | 4.0285(1) | 4.0216(1) | 4.0043(1) | 3.9791(2) |
| *c* (Å) | 8.7416(2) | 8.7251(3) | 8.6909(2) | 8.6359(6) |
| *z* La/Y | 0.1410(2) | 0.1408(2) | 0.1401(2) | 0.1388(5) |
| *z* As | 0.6511(4) | 0.6514(3) | 0.6532(3) | 0.6554(8) |
| (La,Y)-As (Å) | 3.379(2) | 3.373(2) | 3.353(2) | 3.328(4) |
| (La,Y)-O (Å) | 2.361(1) | 2.356(1) | 2.343(1) | 2.323(2) |
| Fe-As (Å) | 2.409(2) | 2.406(2) | 2.404(1) | 2.400(4) |
| As-Fe-As (deg) | 107.5(1) | 107.5(1) | 107.9(1) | 108.2(3) |
| $R_F$ | 3.67 | 2.82 | 3.37 | 5.05 |
| $R_B$ | 4.67 | 3.99 | 4.29 | 6.47 |
| BVS La,Y (vu) | 3.10 | 3.02 | 3.01 | 2.90 |
| BVS O (vu) | 2.30 | 2.24 | 2.22 | 2.14 |
| BVS Fe (vu) | 2.13 | 2.14 | 2.15 | 2.18 |
| BVS As (vu) | 2.93 | 2.92 | 2.94 | 2.94 |

Table 2: Structural, magnetic and resistive transition temperature of $(La_{1-y}Y_y)FeAsO$.

|  | $T_{T-O}$ (K) | | $T_{SDW}$ (K) | $T_{drop}$ (K) |
|---|---|---|---|---|
|  | initial | final | | |
| **y = 0.10** | 153(2) | 126(2) | 142(2) | 142(1) |
| **y = 0.20** | 142(2) | 119(2) | 137(2) | 134(1) |
| **y = 0.30** | 138(2) | 115(2) | 129(2) | 128(1) |

Table 3: Structural data at 10 K for (La$_{1-y}$Y$_y$)FeAsO samples obtained by Rietveld refinement of NPD data (space group no. 67 - *Cmme*;* La,Y at 4$g$; Fe at 4$b$; As at 4$g$; O at 4$a$).

|  | y = 0.10 | y = 0.20 | y = 0.30 |
|---|---|---|---|
| $a$ (Å) | 5.6727(1) | 5.6644(1) | 5.6391(1) |
| $b$ (Å) | 5.7004(1) | 5.6933(1) | 5.6672(1) |
| $c$ (Å) | 8.7179(1) | 8.7073(2) | 8.6725(2) |
| $z$ La/Y | 0.1422(2) | 0.1415(3) | 0.1395(3) |
| $z$ As | 0.6501(3) | 0.6505(4) | 0.6519(4) |
| Fe-As (Å) | 2.399(1) | 2.398(2) | 2.394(2) |
| As-Fe-As (deg) | 107.5(1) | 107.6(1) | 107.8(1) |
| | 107.1(1) | 107.2(1) | 107.4(1) |
| $R_F$ | 3.52 | 3.40 | 3.95 |
| $R_B$ | 4.34 | 4.66 | 5.23 |

* Note that the space group no. 67 was previously referred to as *Cmma*.

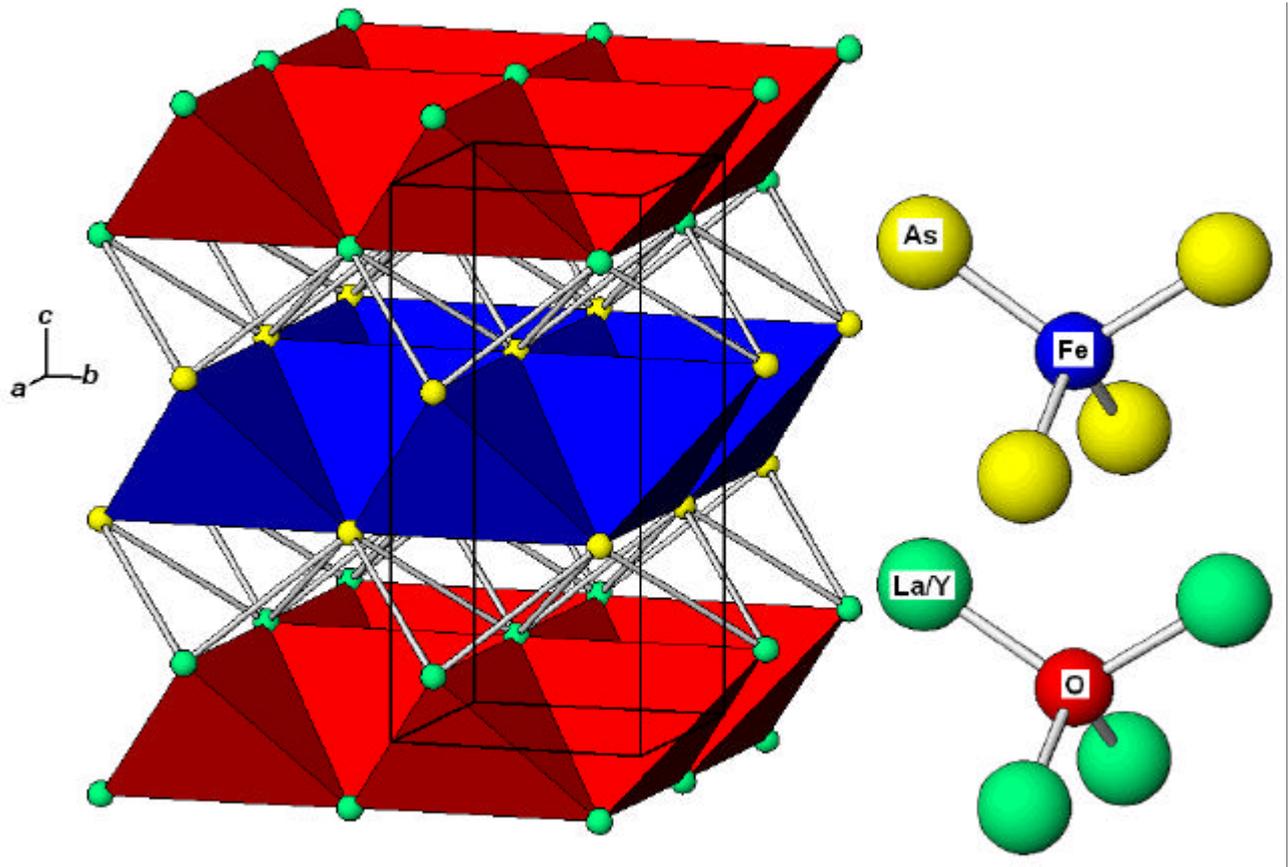

Figure 1 (Color online): Crystal structure of $(La_{1-y}Y_y)FeAsO$ showing the stacking along the $c$ axis of the two kinds of edge-sharing tetrahedral layers, the former centred by Fe with As at corners, the latter by O with La/Y at corners; the layers are linked by (La/Y)-As bonds.

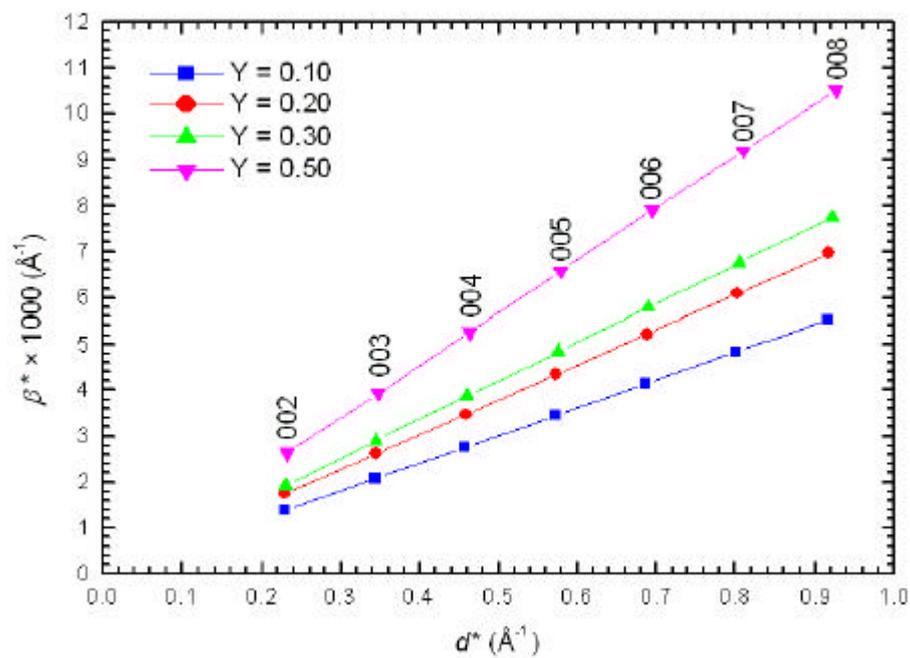

Figure 2 (Color online): Indexed Williamson-Hall plots for the family of 00$l$ peaks of (La$_{1-y}$Y$_y$)FeAsO samples.

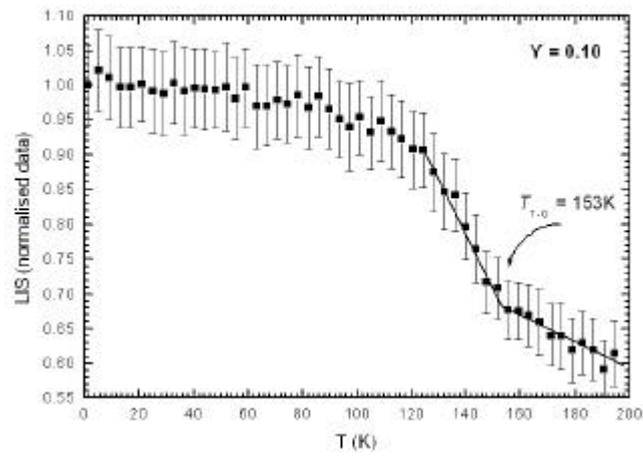

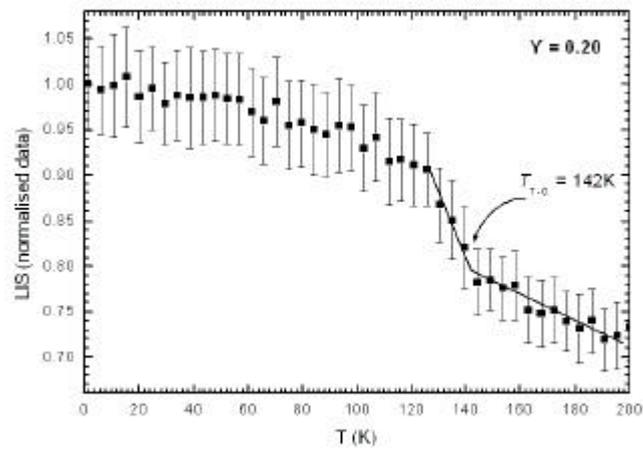

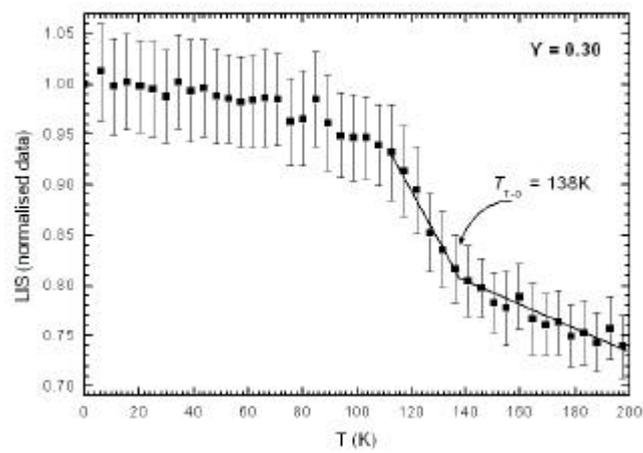

Figure 3: Lorentzian isotropic strain (normalized data) contribution to the 112 diffraction peak of $(La_{1-y}Y_y)FeAsO$ samples as a function of temperature (D1B data); lines are guide to eye.

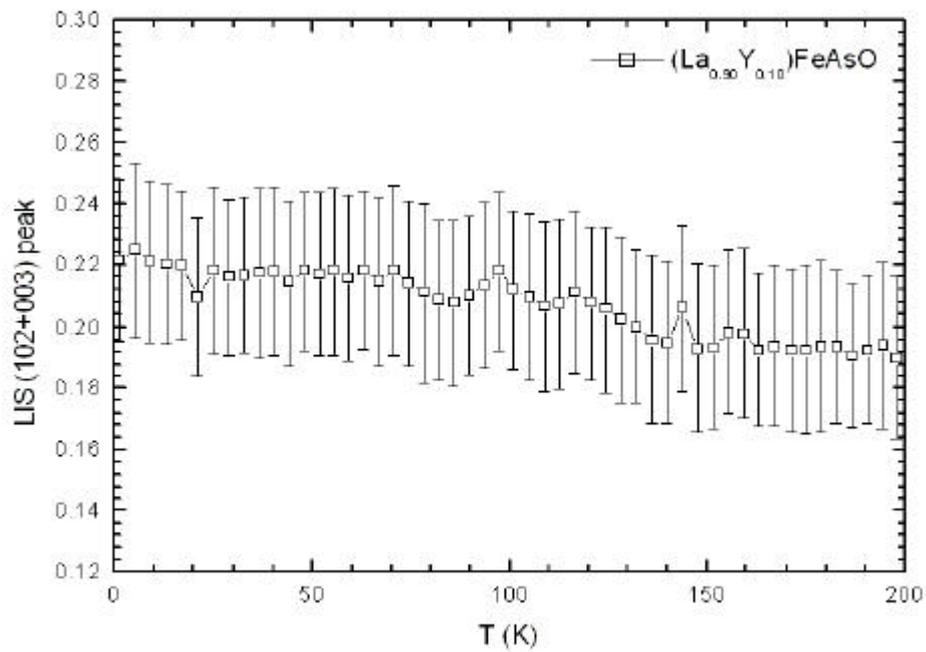

Figure 4: Lorentzian isotropic strain (normalized data) contribution to the to the (102+003) diffraction peak of $(La_{0.90}Y_{0.10})FeAsO$ as a function of temperature (D1B data).

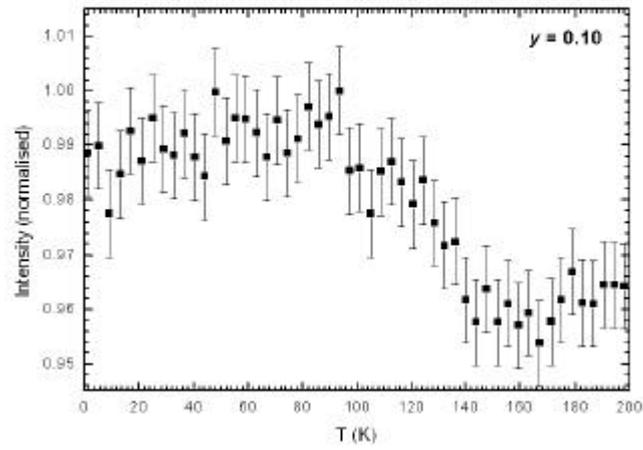

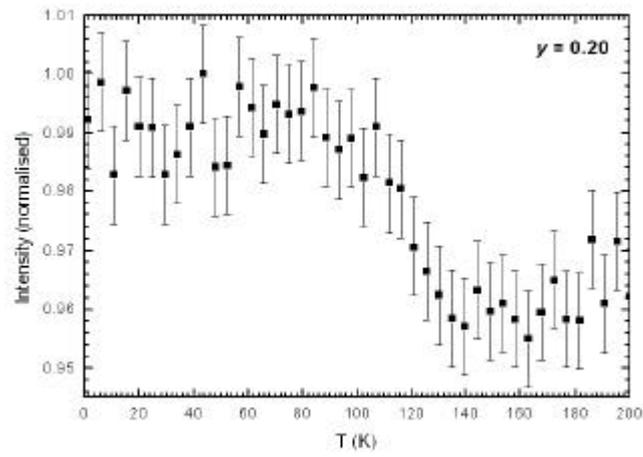

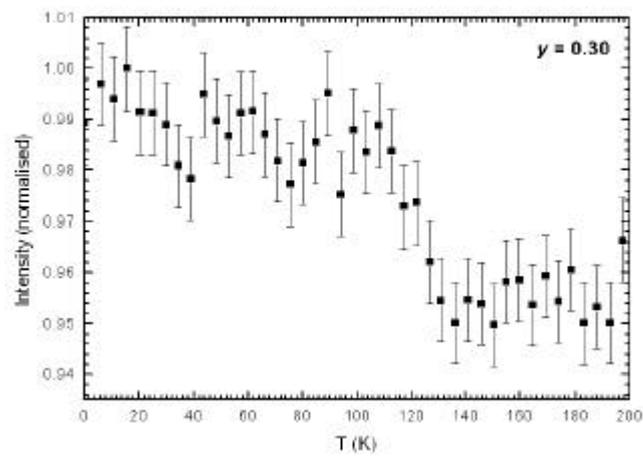

Figure 5: Dependence of the neutron scattering intensity at $d \sim 4.01$ Å on temperature for the $(La_{1-y}Y_y)FeAsO$ samples (normalised data); in Table 2 are reported the corresponding $T_{SDW}$.

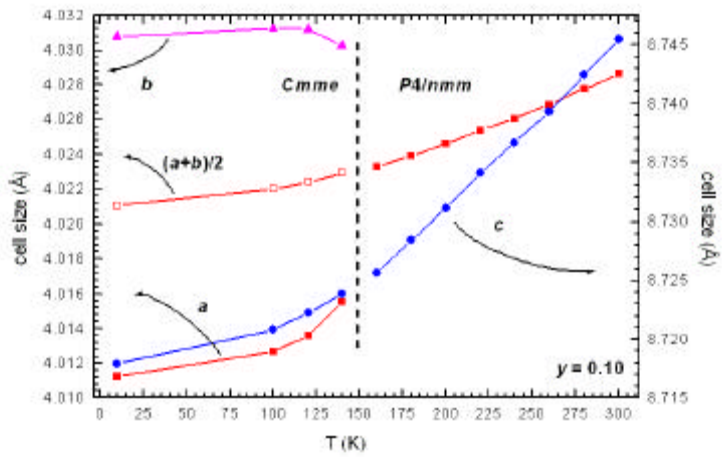
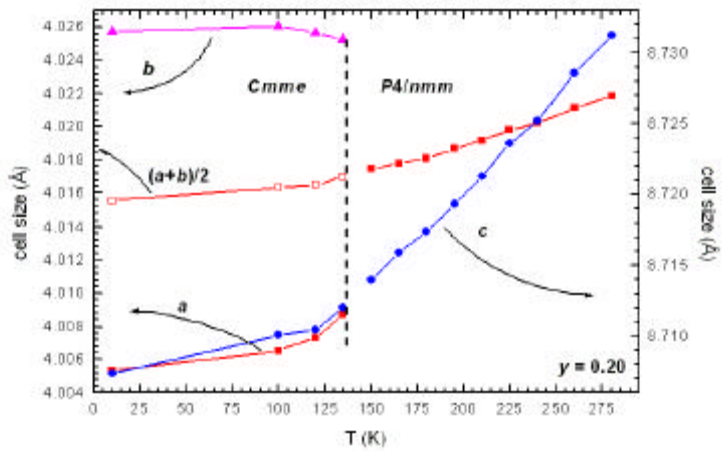
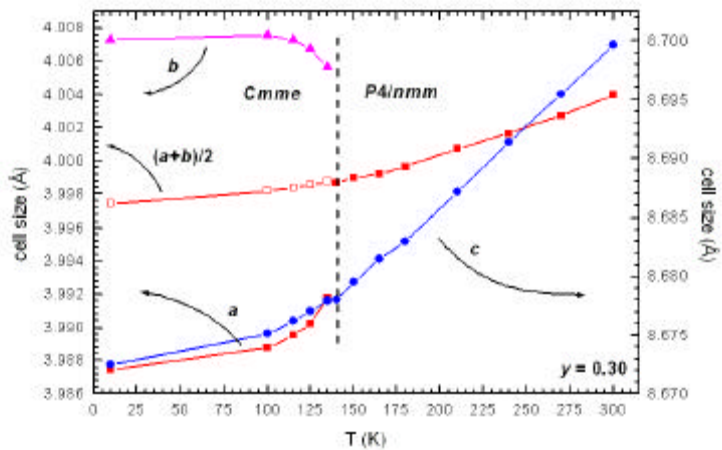

Figure 6 (Color online): Dependence of the cell parameters on $T$ in $(La_{1-y}Y_y)FeAsO$ samples (orthorhombic $a$ and $b$ are divided by $\sqrt{2}$); the average value of orthorhombic $a$ and $b$ parameters (empty squares) follows the behaviour observed for tetragonal cell edge $a$ (full squares).

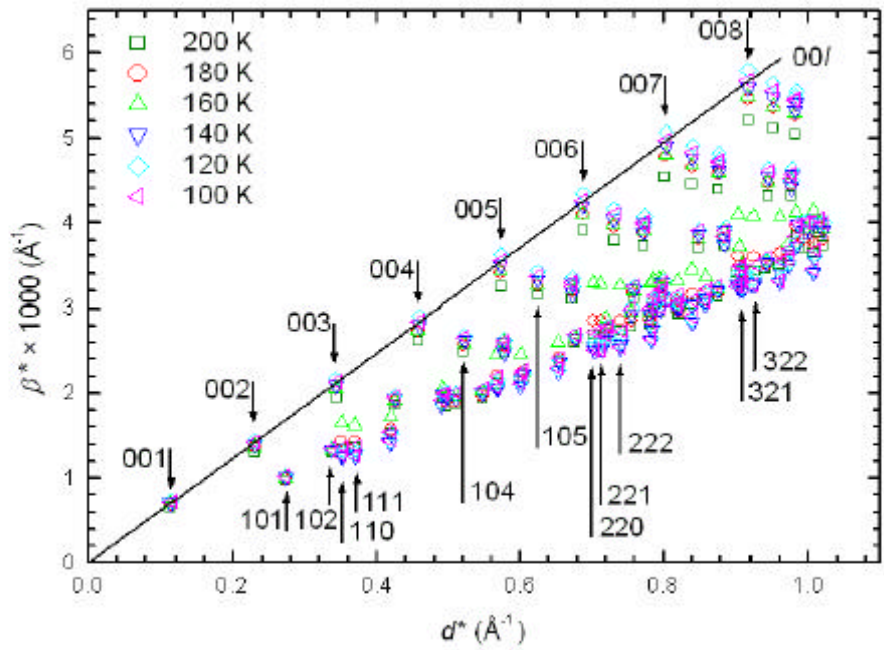

Figure 7 (Color online): Superposition of Williamson-Hall plots, showing the evolution of the strain broadening between 100 K and 200 K in $(La_{0.90}Y_{0.10})FeAsO$; selected peaks are indexed.

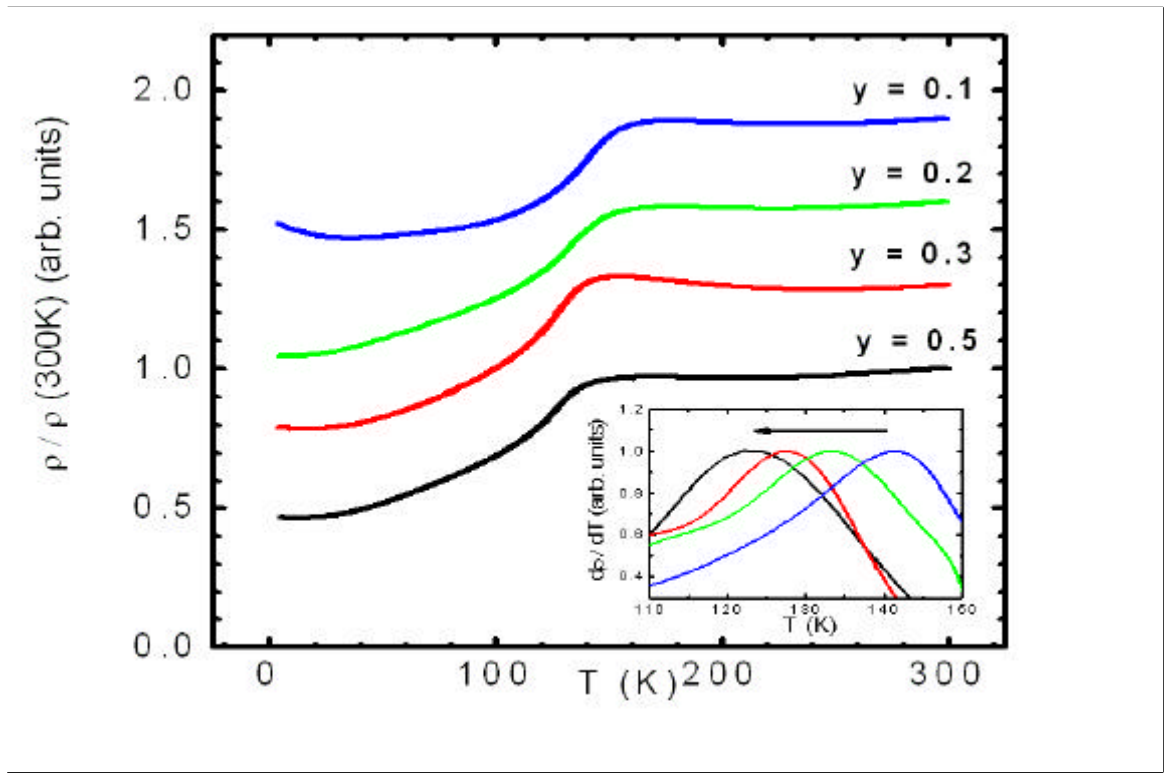

Figure 8 (Color online): $(La_{1-y}Y_y)FeAsO$ normalized resistivity versus temperature behaviour; the inset shows the shift towards low temperature with the increase of the yttrium content of the maximum d*r*/d*T* (normalized).

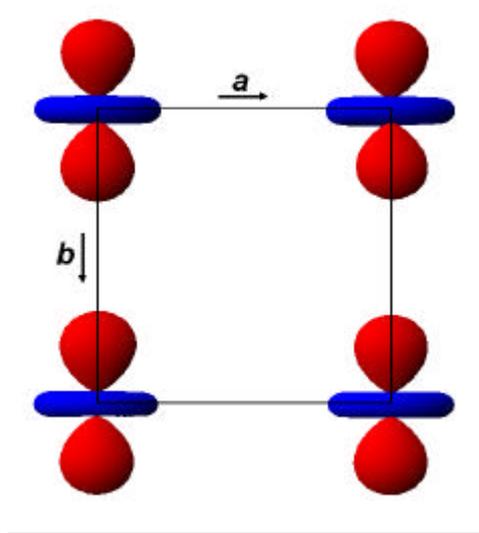

Figure 9 (color online): Alignment of the Fe $3d_{z^2}$ orbitals in the Fe plane along the *b* axis.